\documentclass[
 eqsecnum,aps,prd,nofootinbib]{revtex4}
\usepackage[dvips]{graphicx}
\usepackage{amsmath}
\usepackage{enumerate}
\usepackage{mathrsfs}
\usepackage{amsfonts}

\def\be{\begin{equation}}
\def\ee{\end{equation}}
\def\bea{\begin{eqnarray}}
\def\eea{\end{eqnarray}}
\def\aux { [ \! ]   }
\def\pip { |   }

\begin{document}
%\initfloatingfigs
\title {Solving the Problem of Time in Mini-superspace: \\
Measurement of Dirac Observables}

\author{Donald Marolf\footnote{\tt marolf@physics.ucsb.edu}}

\affiliation{Physics Department, UCSB, Santa Barbara, CA 93106, USA}

\begin{abstract}
One solution to the so-called problem of time is to construct certain Dirac observables, sometimes called evolving constants of motion.  There has been some discussion in the literature about the interpretation of such observables, and in particular whether single Dirac observables can be measured.  Here we clarify the situation by describing a class of interactions that can be said to implement measurements of such observables.   Along the way, we describe a useful notion of perturbation theory for the rigging map $\eta$ of group averaging (sometimes loosely called the physical state ``projector"), which maps states from the auxiliary Hilbert space to the physical Hilbert space.
\end{abstract}

\maketitle

\section{Introduction}
In a universe without a boundary, the diffeomorphism invariance of gravity has two immediate consequences:  First, that the Hamiltonian can be expressed as a sum of constraints, and second, that any observable must commute with these constraints.    It follows that all observables must be time-independent,  and issues surrounding these features are often collectively referred to as ``the problem of time in quantum gravity;" see e.g. \cite{KKCI} for a classic review.  At the classical level, it is of course well understood that formally time-independent observables can nevertheless capture interesting physical information in terms of the {\it relations} between various degrees of freedom.  E.g., we can ask about the average value of the electric field in the region of spacetime occupied by some observer during the time that his/her watch reads between 11:59pm and midnight.    Because the temporal information refers to a physical system (the watch reading), and not to a background notion of time, the corresponding observable does indeed commute with all constraints.

There is by now an extensive literature addressing such relational observables.  The basic idea can be traced back to Einstein and beyond, and of course such ideas are at the root of many gauge-fixing methods.  Concrete methods for constructing relational observables in a manifestly gauge-invariant form by integrating over the spacetime were described by DeWitt \cite{DeWitt}, who also discussed their properties at the semi-classical level. In \cite{evolve1,evolve2,evolve3,evolve4},  the useful term ``evolving constants of motion'' was introduced for observables defined by physical clock and position readings and a number of explicit quantum examples were described.  The observables defined by DeWitt's integration method (and on which we will focus) were then studied at the quantum level in \cite{QORD,BIX,BDT} for mini-superspace models and in \cite{GMH,GMdS} at the level of quantum field theory.  Evolving constants of motion have also been studied in perturbation theory \cite{pert1,pert2,pert3,pert4,pert5}.

While the above references contain various remarks addressing the interpretation of evolving constants,  concerns regarding the degree to which such observables are truly ``physical'' continue to appear in the literature.  In particular, it has been suggested recently \cite{GPPT,FH} that a single relational observable ${\cal O}$ may not have a good physical interpretation, even if ${\cal O}$ is an evolving constant of motion.  At the heart of this discussion (at least in \cite{GPPT}) appears to be a suspicion that such evolving constants cannot actually be measured.   Our purpose here is to clarify such issues by showing that, at least in certain cases, the addition of appropriate interactions {\it can} be said to lead to measurements of such observables.

We intend our discussion to be a relational analogue of von Neumann's classic analysis \cite{vonN}, and indeed there are many similarities.  For definiteness, we consider systems with finite numbers of degrees of freedom which enjoy a time-reparametrization invariance.  Though this excludes any treatment of gravity as a field theory, it does allow us to treat so-called mini-superspace models of quantum gravity, and it is in this class of model that \cite{GPPT,FH} raised objections are precisely of this finite-dimensional type.  Related issues for infinite-dimensional (quantum field theory) systems will be studied in \cite{GMtoappear}, extending the work of \cite{GMH}.

For simplicity, we focus on models described by a Hamiltonian constraint of the form
\begin{equation}
\label{HC0}
0 = H_0 = \frac{1}{2m_c} P_c^2  + \frac{1}{2m_D} P_D^2 + \tilde H_0,
\end{equation}
where the subscript ${}_0$ means that this constraint describes the unperturbed system, before we add the interaction that will lead to our measurement.  Here we imagine that the classical phase space contains two distinguished degrees of freedom $c,D$ and their conjugate momenta $P_c,P_D$, such that $\tilde H_0$ commutes with (and is independent of) $c,D,P_c,P_D$. As a result, the system enjoys two commuting global symmetries associated with translations of $c,D$.    Both $c,P_c$ and $D,P_D$ will play distinguished roles in our measurement setup:  $D,P_D$ will play the role of detector degrees of freedom which will store information about the desired Dirac observable, and $c,P_c$ will play the role of clock degrees of freedom.  The clock allows us both to define the particular Dirac observable of interest (roughly speaking, it is the value of some other degrees of freedom when the clock $c$ reads a certain value) and to construct the interaction that measures it.

We begin with a brief review of formalism in section \ref{formal}, describing both the group averaging approach to quantum constrained systems and the particular class of observables $[B]_{c= \tau}$ that we have in mind which are defined in relation to the clock $c$.  We then briefly describe perturbation theory in this formalism in section \ref{pert}, and in particular the deformation of dynamics induced by adding interaction terms to $H_0$ which are localized with respect to the clock $c$ in a sense explained there.   We also discuss how our deformation is related to the use of retarded or advanced boundary conditions.  This then sets the stage for our measurement theory discussion in section \ref{meas}, where we show than an appropriate such interaction with coupling constant $g$ leads to measurements of $[B]_{c= \tau}$ in the limit of small $g$.  We also argue that these measurements can become arbitarily accurate in the limit of large clock momentum $P_c$.  Finally, we close with a few comments in section \ref{disc}.

\section{Relational Observables and the Physical Hilbert space}
\label{formal}

We are interested in systems defined by a Hamiltonian constraint of the form (\ref{HC0}).  We take the system to be quantized via Dirac Quantization \cite{Dirac}, using in particular the method of Refined Algebraic Quantization via group averaging\footnote{\label{f1}See \cite{ALMMT} for definitions in the language used here.  Group averaging and related techniques were originally introduced in \cite{AH2}, \cite{KL}, \cite{QORD} from various perspectives. See also \cite{single,GM1,GM2,DG,WhereAreWe,GP,OS1,OS2,JLAM03,JLSum,JLAM05} for a number of useful results, and \cite{SumOver,Projector} for discussions in the context of loop quantum gravity.}.  Though we refer to the literature of footnote \ref{f1}  for historical comments and details, we briefly remind the reader of the main ideas associated with this quantization scheme and with the construction of interesting Dirac observables.

The key point is to interpret the constraint operator $H_0$ of (\ref{HC0}) as a self-adjoint operator on some auxiliary Hilbert space ${\cal H}_{aux}$.  Given states $ \aux \psi \rangle$ in an appropriate dense subspace $\Phi$ of ${\cal H}_{aux}$, one defines associated physical states

\begin{equation}
\label{eta0}
\pip \Psi \rangle := \eta_0 \aux \psi \rangle, \ \ {\rm where}  \ \ \eta_0 = \int d \lambda e^{i H_0 \lambda}.
\end{equation}
Because the expression that defines $\eta_0$ is an integral over the group parameter $\lambda$, this form  of $\eta$ is known as the ``group averaging map."  It has the important properties that $\eta \aux \psi \rangle$ solves the constraints ($H_0 \eta_0 =0$) and, furthermore, that any gauge invariant operator on ${\cal H}_{aux}$ (which must necessarily commute with the constraint $H_0$), commutes with $\eta$ (i.e., for such ${\cal O}$ we have $\eta {\cal O} = {\cal O} \eta$).

Any  map with such properties (and for which (\ref{pip}) below is Hermitian and positive-definite) is called a ``rigging map" \cite{ALMMT}.  It is useful to use the rigging map to introduce an inner product on states of the form (\ref{eta0}) via

\begin{equation}
\label{pip}
 \langle \Psi_2 \pip \Psi_1 \rangle : =  \langle \psi_1 \aux \eta \aux \psi_2 \rangle = \int d \lambda \langle \psi_1 \aux e^{i \lambda H_0} \aux \psi_2 \rangle.
\end{equation}
For appropriate choices of subspace $\Phi \subset {\cal H}_{aux}$, this expression is finite and positive definite when zero lies in the continuous spectrum of $H_0$.  Completing the image of $\eta$ in this inner product then yields the physical Hilbert space ${\cal H}_{phys}$.

Having described the physical Hilbert space, we now wish to construct some interesting Dirac observables, by which we mean operators ${\cal O}$ on ${\cal H}_{aux}$  which commute with $H_0$.  Because they commute with $\eta$, any such observable can then be promoted to an operator on ${\cal H}_{phys}$.  While it is typically rather difficult to find explicit combinations of the coordinates and momenta which commute with $H_0$, it was noted in \cite{QORD} that one may express a large class of such observables using an integral representation analogous to that used to define the rigging map $\eta$ above.  One simply picks some fixed operator $A$ on ${\cal H}_{phys}$ and observes that the expression
\begin{equation}
\label{obs}
{\cal O} : = \int d \lambda \ e^{iH_0 \lambda} \ A \  e^{-iH_0 \lambda}
\end{equation}
formally commutes with $H_0$ due to the translation-invariance of the measure $d\lambda$.  As a result, (\ref{obs}) defines an observable on ${\cal H}_{phys}$ so long as the integral converges weakly on the dense subspace $\Phi \subset {\cal H}_{aux}$.  It is natural to refer to observables of the form (\ref{obs}) as ``single-integral observables; " see \cite{GMH,GMdS} for quantum discussions of field theoretic generalizations.  For later purposes, we note that the definition of the physical inner product (\ref{pip}) allows one to write matrix elements of such ${\cal O}$ between physical states $\pip \Psi_{1,2} \rangle$ in terms of the corresponding auxiliary states $\aux \psi_{1,2} \rangle$ as
\begin{equation}
\label{pme}
\langle \Psi_1 \pip {\cal O} \pip \Psi_2 \rangle = \int d\lambda_1 d \lambda_2 \langle \psi_1 \aux e^{iH_0 \lambda_1} A e^{-iH_0 \lambda_2} \aux \psi_2 \rangle;
\end{equation}
in particular, only two group-averaging integrals appear in this expression.

Expressions of the form (\ref{obs}) appear straightforward to interpret.  The conjugation by $e^{iH_0 \lambda}$ may be thought of as translating the operator $A$ in the parameter time $\lambda$.  Thus, (\ref{obs}) is essentially an integral over all parameter times; i.e., over the 0+1 spacetime which defines our system.  As such, (\ref{obs}) is essentially a quantum version of DeWitt's construction \cite{DeWitt}.  It is no surprise that such integrals define reparametrization-invariant observables\footnote{To write (\ref{obs}) in an explicitly reparametrization-invariant form requires the introduction of a lapse function $N(\lambda)$ as described in \cite{QORD}.}.  Furthermore, one way to get good convergence properties is to take $A$ to be localized in some particular region of phase space.  In particular, for any $\tau \in {\mathbb R}$, we will be interested in operators of the form $A = BF_\tau$, where $B$ is any operator that commutes with $c, P_C$ and $F_\tau$ is constructed from the clock degrees of freedom as follows:  First, choose some smooth function $f(c)$ which approximates a step function at $c=0$.  Specifically, we require $f(-\infty)=0$ and $f(+\infty) =1$, and we require the derivative of $f$ to vanish outside some compact interval.   Given this $f(c)$, we define

\begin{eqnarray}
\label{Ftau}
F_\tau &=& \frac{1}{2} \{ {\rm sign} P_c, \frac{\partial}{\partial \lambda} f(c-\tau) \} \cr
&=& i \frac{1}{2} \{ {\rm sign} P_c,  [\frac{1}{2m_c} P_c^2 , f(c-\tau) ] \}
=    \frac{1}{2m_c } \{ {\rm sign} P_c,  \{ f'(c-\tau), P_c \} \},
\end{eqnarray}
where $\{,\}$ denotes the anti-commutator, ${\rm sign} P_c$ denotes the sign of $P_c$, and the action of $\frac{\partial}{\partial \lambda}$ is defined using the Heisenberg equation of motion $\frac{\partial}{\partial \lambda} f := i [H_0, f].$    At the classical level, if we take $f(c)$ to be an exact step function, it is clear that single integral observables of the particular form
\begin{equation}
\label{Bct}
[B]_{c= \tau} := \int d \lambda e^{iH_0 \lambda} B F_\tau e^{-iH_0 \lambda}
\end{equation} give precisely the value of $B$ at the particular event where $c = \tau$.  Below we will focus on the quantum versions of the relational single-integral observables $[B]_{c= \tau}$ given by (\ref{Bct}).  See \cite{QORD,BIX,BDT} for studies of these observables in various models.  In particular, \cite{QORD} discusses conditions under which related observables are self-adjoint for simple models.  Self-adjointness appears most likely when $B$ is a bounded operator, so that the classical analogue of (\ref{Bct}) is a bounded function on phase space.

\section{Perturbations and the Physical Hilbert space}
\label{pert}

Our discussion thus far has focused on Hamiltonian constraints of the particular form (\ref{HC0}), in which the detector degrees of freedom ($D,P_D$) are coupled to other degrees of freedom only through the constraint.  In particular, the detector momentum $P_D$ is a conserved quantity.  These systems contain interesting observables of the form $[B]_{c= \tau} $ which we might like to measure.   Of course, a dynamical measurement requires some physical interaction between the detector degrees of freedom $D,P_D$ and the quantity to be measured.  No such interaction is present in (\ref{HC0}).  Instead, our task will be to modify the Hamiltonian constraint (\ref{HC0}) by introducing an appropriate coupling (with some coupling constant $g$), and to show that this leads to a measurement of $[B]_{c= \tau} $ in the limit of small $g$.  As a result, and as is typical in discussions of measurement theory, our argument will perturbative in nature.

In perturbation theory for constrained systems, one must take into account the fact that the  constraint itself, and thus the very definition of the physical Hilbert space,  will depend on the coupling $g$.  As a result, if one attempts to work directly with physical Hilbert spaces, it appears to make little sense to think of a state at finite $g$ as a perturbation of a $g=0$ state.  In particular, it is not clear in what space such a perturbation could possibly live.

Nevertheless, by introducing the auxiliary Hilbert space ${\cal H}_{aux}$, refined algebraic quantization (and, in particular, group averaging) can be used to set up a well-defined perturbation theory.  The point is that, at least in our finite-dimensional context, physical states for all $g$ may be defined as linear functions on the same dense subspace $\Phi$ of  ${\cal H}_{aux}$.  Thus, any perturbation of the state is also naturally thought of as a linear function on $\Phi$.  In somewhat more physical language one might say that both the $g=0$ physical state and the perturbation are ``non-normalizeable'' states in the same space ${\cal H}_{aux}$.

In more concrete terms, consider some auxiliary state $\aux \psi \rangle$ chosen to be independent of $g$.  Then for each $g$ we may consider the physical state
\begin{equation}
\label{gphys}
\pip \Psi_g \rangle := \eta_g \aux \psi \rangle :=  \int d\lambda e^{iH_g \lambda} \aux \psi \rangle,
\end{equation}
where the subscripts $g$ denote the objects at coupling strength $g$. So far, this is merely a recipe for obtaining a one-parameter family of states $\pip \Psi_g \rangle$ which reduce to $\pip \Psi_0 \rangle$ in the $g \rightarrow 0$ limit; we have not yet provided any particular physical interpretation for this family.  We will return to this issue below, after completing the formulation of our perturbation theory.  At that point we will be able to state conditions under which (\ref{gphys}) corresponds to an analogue of choosing retarded (or advanced) boundary conditions for our relational system.

First, however, we must state the sense in which we will expand (\ref{gphys}) in powers of $g$. The trick, of course, is to formulate our  perturbation theory in a way that yields finite answers at each order in $g$.   Since $\eta_g$ is a somewhat singular object (it may be thought of as roughly the Dirac delta function $\delta(H_g)$ of the operator $H_g$), we find it convenient to work instead with the unitary operator $e^{iH_g \lambda}$ appearing in (\ref{gphys}).   In particular, let $H_g = H_0 + g \Delta H$ and recall that $e^{i(H_0 + g \Delta H) \lambda} = \lim_{N \rightarrow \infty} [1 + i (H_0 + g \Delta H) \frac{\lambda}{N}]^N.$  Expanding this result in powers of $g$ and then taking the limit yields the useful expression

\begin{eqnarray}
\label{Uexp}
e^{i H_g \lambda} &=& e^{i H_0 \lambda} + i g \int_{{\lambda_1+ \lambda_2 = \lambda}  \atop { \lambda_1, \lambda_2 \ge 0}} d \lambda_1 e^{i H_0 \lambda_1} \Delta H e^{i H_0 \lambda_2}  + \dots   \cr &+&  (i g)^k \int_{{ \sum_{j=1}^k \lambda_j = \lambda}  \atop { \lambda_j  \ge 0}} \left( \prod_{i \le k-1} d \lambda_i \right)  e^{i H_0 \lambda_1} \Delta H e^{i H_0 \lambda_2} \dots    e^{i H_0 \lambda_{k-1}} \Delta H e^{i H_0 \lambda_k}  + \dots  \  {\rm for} \ {\lambda \ge 0}, \  \ \ \ \ \ \
\end{eqnarray}
and a corresponding expression involving $\lambda_i \le 0$ for $\lambda \le 0$ which is most easily obtained by taking the Hermitian conjugate of (\ref{Uexp}).
In (\ref{Uexp}), each term simply has $k-1$ insertions of $i g \Delta H$ occupying all possible positions subject to the restriction that any two neighboring insertions ($j, j+1$) are separated by a translation involving some parameter time $\lambda_j$ of the same sign as the total parameter time $\lambda$.  This coincides with the perturbation theory used in \cite{pert1,pert2,pert3,SumOver} when $H_0=0$.  See \cite{pert4,pert5} for related treatments of (classical) perturbation theory with $H_0 \neq 0$.

It is also useful to rewrite each term above by thinking of each operator $\Delta H$ as being the $\lambda=0$ value of some Heisenberg-picture operator in the free theory.  I.e., we define $\Delta H_0(\lambda) = e^{iH_0 \lambda} \Delta H e^{-iH_0 \lambda}.$  Introducing $\tilde \lambda_k = \sum_{i=1}^k \lambda_k$ then yields (again, for $\lambda \ge 0$)
\begin{eqnarray}
\label{Uexp2}
 e^{i H_g \lambda} &=& e^{i H_0 \lambda} + i g \int_{\lambda \ge \tilde \lambda_1 \ge 0} d \tilde  \lambda_1 \Delta H_0(\tilde  \lambda_1)  e^{i H_0 \lambda}  + \dots   \cr &+&  (i g)^k \int_{\lambda \ge \tilde \lambda_k \ge \dots  \tilde \lambda_1 \ge 0} \left( \prod_{i \le k-1} d \tilde  \lambda_i \right)  \Delta H_0 (\tilde  \lambda_1)   \Delta H_0 (\tilde  \lambda_2)  \dots  \Delta H_0 (\tilde \lambda_k) e^{i H_0 \lambda}   + \dots  \  {\rm for} \ {\lambda \ge 0}. \ \ \ \ \ \ \ \
\end{eqnarray}
In this form one recognizes (\ref{Uexp2}) as the standard interaction picture perturbation theory familiar from quantum field theory, though we have expressed this result in a slightly different language.
In particular, expression (\ref{Uexp2}) can be used to compute the result of time-evolving any operator $A$ with $e^{iH_g \lambda}$ (for $\lambda > 0$) :
\begin{eqnarray}
\label{conjexp}
e^{iH_g \lambda} A e^{-iH_g \lambda} &=& A_0(\lambda) + ig \int_{\lambda \ge \tilde \lambda_1 \ge 0 } d \tilde \lambda_1
[\Delta H_0(\lambda_1), A_0(\lambda)] \cr &+& (ig)^2 \int_{\lambda \ge \tilde \lambda_1  \ge \tilde \lambda_2 \ge 0} d \tilde \lambda_1 d \tilde \lambda_2
[\Delta H_0(\tilde \lambda_2), [\Delta H_0(\tilde \lambda_1), A_0(\lambda)] ] + \dots
\end{eqnarray}
where we have again defined $A_0(\lambda) =
e^{iH_0 \lambda} A e^{-iH_0 \lambda}$ to be the free evolution.

Inserting either (\ref{Uexp}) or (\ref{Uexp2}) into (\ref{gphys}) gives a perturbation series for $\pip \Psi \rangle_g$. We can now use this perturbation series to say something about the physical interpretation of the states $\pip \Psi \rangle_g$ for systems where $H_0$ is of the form (\ref{HC0}) and where the perturbation $\Delta H$ commutes with $c$ and is localized near some clock reading $\tau_0$; e.g., we might have $\Delta H = F_{\tau_0} B D$ where $F_{\tau_0}$ is again defined by (\ref{Ftau}) and thus is peaked near $\tau_0$.  This choice of $\Delta H$ will prove useful in section \ref{meas} below.  Suppose we choose the auxiliary state $\aux \psi \rangle$ to be supported on large positive clock momenta $P_c > 0$ and to be peaked near some value $\tilde \tau$ of $c$ which satisfies $\tilde \tau < \tau_0$.  Note that since we consider an auxiliary state, these criteria make sense even though $c$ is a gauge-dependent operator (as is $P_c$ for $g \neq 0$). Then, at least in the limit $\tilde \tau \ll \tau_0$, the choice (\ref{gphys}) can be thought of as imposing a sort of relational retarded boundary conditions.  By this we mean that, at least in perturbation theory, any family of operators of the form

\begin{equation}
\label{Bgct}
[B_k]^g_{c= \tau} := \int d \lambda e^{iH_g \lambda} B_k  F_\tau e^{-iH_g \lambda}
\end{equation}

satisfies
\begin{equation}
\label{ret}
\langle \Psi_g \pip [B_1]^g_{c= \tau_1} [B_2]^g_{c= \tau_2}  \dots [B_N]^g_{c= \tau_N} \pip \Psi_g \rangle \rightarrow \langle \Psi_0 \pip [B_1]^0_{c= \tau_1} [B_2]^0_{c= \tau_2}  \dots [B_N]^0_{c= \tau_N} \pip \Psi_0 \rangle
\end{equation}
in the limit $\{\tau_i \}, \tilde \tau \ll \tau_0$.  Here as usual we require the $B_i$ to commute with $c, P_c$ .

The result (\ref{ret}) is easily derived by using (\ref{Uexp}) in both (\ref{gphys}) and (\ref{Bgct}).  It is clear that the right-hand-side of (\ref{ret}) corresponds to the zero-order term in $g$.  To see why the other terms vanish, consider the first order correction for the simple case $N=1$. Using the $g \neq0$ analogue of (\ref{pme}), this correction is:

\begin{eqnarray}
\label{first}
i g  \langle \psi \aux  \int d\lambda_3 \int_{{\rm sign}(\lambda_1) = {\rm sign}(\lambda_2) } d \lambda_1 d\lambda_2 \ {\rm sign}(\lambda_1)
&\Bigl(& e^{i H_0 \lambda_3}
B F_\tau
e^{i H_0 \lambda_1} \Delta H e^{i H_0 \lambda_2} \cr
&+&
e^{i H_0 \lambda_1} \Delta H e^{i H_0 \lambda_2}
B F_\tau
e^{i H_0 \lambda_3}
 \Bigr)
 \aux \psi \rangle.
\end{eqnarray}
But recall that $F_\tau$ is built via (\ref{Ftau}) from $f'(c-\tau)$, which has support only near $c=\tau$.  In the same way, $\Delta H$ is peaked near $c = \tau_0$, and $\aux \psi \rangle$ is peaked near $c= \tilde \tau$.   Thus, in the limit $\tau_0 \gg \tilde \tau, \tau$ the first term in  (\ref{first}) is very small unless $\lambda_2$ has the right value to locate the peak of $\langle c \aux e^{iH_0 \lambda} \aux \psi \rangle$ at $\tau_0$; note that this will largely fix $\lambda_2$ to some particular (postive) value.  On the other hand, the first term is also small unless $\lambda_1$ has the right value to move the peak from $\tau_0$ to $\tau$, requiring $\lambda_1 < 0$ since $\tau_0 \gg \tau$.  But only contributions with ${\rm sign}(\lambda_1) = {\rm sign}(\lambda_2)$ are allowed, so these conditions cannot be satisfied simultaneously.  As a result, the first term in (\ref{first}) must vanish in the limit $\tau_0 \gg \tilde \tau, \tau$.

The second term in (\ref{first}) behaves similarly, as do the higher order corrections from (\ref{Uexp}). We therefore conclude that (\ref{ret}) holds in perturbation theory, meaning that it holds if we interpret the left-hand side as being determined by the expansion (\ref{Uexp}) to any finite order in $g$.  Note that since all derivatives of $f$ vanish outside of a compact interval, the various factors of $P_c$ pose no problem for the above argument:  they can be commuted out of the way at the price of taking further derivatives of $f$, which behave similarly.  Also, in the limit where $\langle \psi \aux$ is supported  at large positive $P_c$, the factors of ${\rm sign} P_c$ can be neglected.

It is worth commenting on some obvious generalizations of this result.  First, note that (\ref{ret}) also holds in the limit $\{\tau_i \}, \tilde \tau \gg \tau_0$.  Thus, for $\aux \psi \rangle$ peaked near $\tilde \tau > \tau_0$, the states $\pip \Psi \rangle_g$ defined by (\ref{gphys}) may be thought of as satisfying an {\it advanced} boundary condition.  Second, while we have stated result (\ref{ret}) for expectation values, it is clear that an analogous statement also holds for matrix elements of $[B_1]^g_{c= \tau_1} [B_2]^g_{c= \tau_2}  \dots [B_N]^g_{c= \tau_N}$ between two distinct physical states.

\section{Measuring Evolving Constants of Motion}
\label{meas}

Sections \ref{formal} and \ref{pert} have introduced most of the technical results needed to describe measurements of the evolving constants of motion $[B]_{c=\tau}$ for systems with $H_0$ of the form (\ref{HC0}).   What remains is to compute the response of an appropriate detector variable (which we choose to be essentially $P_D$) under the inclusion of an interaction of the form $\Delta H = \Delta H = F_{\tau_0} B D$ (with $F_{\tau_0}$  given as usual by (\ref{Ftau})) as suggested in section \ref{pert}.
We do so in section \ref{DOM} below and argue that the result may be interpreted as a measurement of $[B]_{c=\tau}$.  We then discuss the accuracy of such measurements in section \ref{res}.

\subsection{Detector Observables and the Measurement}
\label{DOM}

The reader will note that $P_D$ is not an observable for $g\neq 0$.  We must therefore state in what sense we will compute the effect of the interaction on $P_D$.  One notion of this effect closely mirrors the usual discussion of measurement theory for systems with a well-defined external time (see e.g. \cite{vonN,DeWitt}) and is given by the difference between $P_D$ evaluated at $\lambda = \pm \infty$ in the interacting theory (finite $g$).   While $P_D$ at any finite $\lambda$ is gauge-dependent, the $\lambda \rightarrow \pm \infty$ limit is gauge invariant because, as $\lambda \rightarrow \pm \infty$, the wavefunction has support only at $c \rightarrow \pm \infty$, effectively turning off the interaction so that the asymptotic evolution in $\lambda$ is generated by the free term $H_0$ (which commutes with $P_D$).  I.e., in this limit the commutator $[P_D, H_g] = g  F_{\tau_0} B$ vanishes weakly.  Thus, it is physically sensible to compute $\Delta P_D := P_{D,g}(\lambda = + \infty) - P_{D,g}(\lambda = - \infty)$, where for any operator $A$ we define $A_g(\lambda) = e^{iH_g \lambda} A e^{-iH_g \lambda} $, using the subscript $g$ to distinguish this notion of evolution from the $g=0$ evolution defined by $A_0(\lambda)$ in section \ref{pert}.  We will show below that $\Delta P_D = -g [B]_{c=\tau} + O(g^2)$, so that our proposed interaction correlates detector degrees of freedom with $[B]_{c=\tau}$ in the limit of small $g$.  Such a situation is naturally described as being a measurement of $[B]_{c=\tau}$.

Before doing so, however, we pause to show that $P_{D,g}(\lambda = \pm \infty)$ in fact agrees with the gauge invariant observable $\lim_{\tau \rightarrow \pm \infty}[P_D]^g_{c = \tau}$, at least in the limit of large clock momentum $P_c$. The difference $[P_D]^g_{c = +\infty} - [P_D]^g_{c = -\infty} $ defines a second notion of the effect of our interaction on $P_D$ which is manifesly relational and gauge invariant.   The equality $P_{D,g}(\pm \infty) =  [P_D]^g_{c = \pm \infty}$ will establish that the result of our measurement is stored in observables of the same general type as the observable measured by our interaction.  It will therefore follow that adding additional interactions to other detectors would allow similar measurements of our apparatus' memory, demonstrating that there is no obstacle to accurately reading out the results of the measurement.

To begin the argument, consider matrix elements of $[P_D]^g_{c = \tau}$ between two physical states $\pip \Psi_{1g,2g} \rangle = \eta_g \aux \psi_{1,2} \rangle$ defined at each $g$ by the auxiliary states
$ \aux \psi_{1,2} \rangle$.  We have

\begin{equation}
\label{PDme}
\langle \Psi_{1g} \pip [P_D]^g_{c = \tau} \pip \Psi_{2g} \rangle =
\langle \psi_1 \aux \int d \lambda   e^{i\lambda H_g} F_{\tau}  e^{-i\lambda H_g} e^{i\lambda H_g} P_D e^{-i\lambda H_g} \ \ \eta_g \aux \psi_2 \rangle,
\end{equation}
where we have merely used the finite $g$ analogue of (\ref{pme}),  the definition of $P_D$,  and the fact that $e^{-i\lambda H_g} e^{i\lambda H_g} =1.$  Expression (\ref{PDme}) can be simplified in the limit where $ \aux \psi_{1,2} \rangle$ are supported at large clock momentum $p_c$.  As described in appendix \ref{freeclock}, the clock degrees of freedom evolve freely in that limit and in particular we have
\begin{equation}
\langle \psi_1 \aux e^{iH_g \lambda} F_\tau e^{-iH_g \lambda}
= \langle \psi_1 \aux  e^{iH_0 \lambda} F_\tau e^{-iH_0 \lambda}  + O(1/p_c).
\end{equation}

Next, we note that  (as in section \ref{pert}) for large positive $\tau$ the factor $\langle \psi_1 \aux  e^{iH_0 \lambda} F_\tau e^{-iH_0 \lambda}$ is very small unless $\lambda$ is very large.  Thus, for large positive (negative) $\tau$, we need only consider
$e^{i\lambda H_g} P_D e^{-i\lambda H_g}$ for large positive (negative) $\lambda$.  As discussed above, the limits $P_{D,g} (\lambda = \pm \infty) = \lim_{\lambda \rightarrow \pm \infty}  e^{i\lambda H_g} P_D e^{-i\lambda H_g}$ are well defined\footnote{This may be explicitly verified in perturbation theory using (\ref{conjexp}).}, and can thus be taken outside the integral over $\lambda$. We find
 \begin{eqnarray}
 \label{PDme2}
\lim_{\tau \rightarrow \pm \infty} \langle \Psi_{1g} \pip [P_D]^g_{c = \tau} \pip \Psi_{2g} \rangle &=&
\langle \psi_1 \aux \left( \int d \lambda   e^{i\lambda H_0} F_{\tau}  e^{-i\lambda H_0} e^{i\lambda H_g} \right) P_{D,g}(\pm \infty)  \ \ \eta_g \aux \psi_2 \rangle \cr
&=&
\langle \psi_1 \aux [\openone]^0_{c = \tau} P_{D,g}(\pm \infty)  \ \ \eta_g \aux \psi_2 \rangle,
 \end{eqnarray}
where $[ \openone]^0_{c= \tau}$ is defined as in (\ref{Bct}) with $B$ chosen to be the unit operator $\openone$; i.e., it is the factor in parentheses on the right-hand-side of the first line of (\ref{PDme2}).

Clearly, we wish to better understand the operator $[\openone]^0_{c = \tau}$, which one notes is built only from clock operators $\tau_0(\lambda), P_{c,0}(\lambda)$ undergoing {\it free} evolution.  In fact, for all $\tau$ we have

\begin{equation}
\label{idt}
[\openone]^0_{c = \tau} = \int d \lambda e^{iH_0 \lambda}  \frac{1}{2} \{ {\rm sign} P_c, \frac{\partial}{\partial \lambda} f(c-\tau) \} \Big|_{\lambda = 0} e^{-iH_0 \lambda}
=   \frac{1}{2} \{ {\rm sign} P_c,  \int d \lambda   \frac{\partial}{\partial \lambda} f(c-\tau) \} ,
\end{equation}
where the final integral is straightforward to perform:
\begin{equation}
\label{intdf}
\int d \lambda   \frac{\partial}{\partial \lambda} f(c-\tau) = f(c-\tau) \Big|_{\lambda = +\infty} -  f(c-\tau) \Big|_{\lambda = -\infty} = {\rm sign} P_c.
\end{equation}
Here, in the last step we have used the fact that, in the free clock system, states with positive $P_c$ move toward $c = +\infty$ (where $f(c-\tau) = 1$) as $\lambda \rightarrow +\infty$, and states with negative $P_c$ move toward $c = -\infty$ (where $f(c-\tau) = 0$) as $\lambda \rightarrow +\infty$ (and vice versa as $\lambda \rightarrow -\infty$). The last equality in (\ref{intdf}) can also be explicitly verified by computing matrix elements.   For the matrix element calculation, we refer the interested reader to the discussion leading up to equation (4.14) of \cite{QORD}.  This discussion performs essentially the same calculation if one sets $A = \openone$ in that reference.
Inserting (\ref{intdf}) into (\ref{idt}) yields
\begin{equation}
\label{1}
[\openone]^0_{c = \tau} = \openone.
\end{equation}
As a result, (\ref{PDme2}) can be written
\begin{equation}
 \label{PDme3}
\lim_{\tau \rightarrow \pm \infty}  [P_D]^g_{c = \tau} =  P_{D,g}(\pm \infty)   + O(1/p_c).
 \end{equation}
 I.e., as claimed, the large $\tau$ limits of the relational observables $ [P_D]^g_{c = \tau}$ are just the limits $ P_{D,g}(\pm \infty)$ of $P_D$ at large parameter times.

Having introduced the detector observables of interest, we now compute the effect of our interaction on these observables.  To do so, we need only expand $\Delta P_D : = lim_{\lambda \rightarrow +\infty} \left( e^{iH_g \lambda} P_{D} e^{-iH_g \lambda} - e^{-iH_g \lambda} P_D e^{iH_g \lambda}\right)$ in powers of $g$.  This computation is readily performed using (\ref{conjexp}) and, using the fact that $P_D$ is constant under the free evolution generated by $H_0$, yields
\begin{eqnarray}
\label{DPD}
\Delta P_D
&=& ig \int d\lambda_1 [\Delta H_0(\lambda_1),P_D]  +O(g^2)  \cr
&=& -g \lim_{\lambda_\pm \rightarrow \pm \infty} \int_{\lambda_-}^{\lambda_+} d\lambda e^{iH_0 \lambda} F_{\tau_0} B e^{-iH_0 \lambda}  + O(g^2) = ig [B]_{c=\tau_0} + O(g^2).
\end{eqnarray}
 Thus, to lowest order in $g$, $\Delta P_D$ is perfectly correlated with $[B]_{c=\tau_0}$.  In analogy with \cite{vonN}, we interpret this as a measurement of $[B]_{c=\tau_0}$ by our detector.

\subsection{High Resolution Measurements}
\label{res}

Equation (\ref{DPD}) is our main result.  However, it remains to analyze the accuracy with which these measurements can be made.  The point is that (\ref{DPD}) correlates
$[B]_{c=\tau_0}$ with $\Delta P_D= P_{D,g}(\infty) - P_{D,g}(-\infty) $.  However, the interpretation of the measurement greatly simplifies if we can treat $P_{D,g}(-\infty)$ as known so that the interaction would correlate $[B]_{c=\tau_0}$ with the final state of the measuring device as described by $P_{D,g}(\infty)$.  This requires choosing a state of the system in which the value of $P_{D,g}(-\infty)$ is sharply peaked.  While this can easily be done, one must ask whether the $O(g^2)$ corrections can be made arbitrarily small in such a sharply peaked momentum state.  The issue is that, as is typically the case for measurements, the conjugate operator $D$ to $P_D$ appears in the $O(g^2)$ correction (see e.g. \cite{BR1,BR2,DeWitt}). Thus, for any fixed interaction with $g \neq 0$, the $O(g^2)$ correction will necessarily become large in the limit in which $P_{D,g}(-\infty)$ is sharply peaked .  If one tries to minimize the total effect of these two sources of error, one must deal with the fact that (say, under our retarded boundary conditions) the uncertainties $\Delta_{unc} D$ and $\Delta_{unc} P_D$ in $D, P_{D,g}(-\infty) = P_{D,0}(-\infty) $ satisfy $(\Delta D)  (\Delta P_D) \ge 1$.  Since rearranging (\ref{DPD}) yields
\begin{equation}
[B]_{c = \tau} + \frac{1}{g} P_D(+\infty) = \frac{1}{g} P_D(-\infty) + g X + O(g^2),
\end{equation}
where $X$ is the second order correction to (\ref{DPD}) linear in $D$ mentioned above,
an uncertainty in $P_{D,g}(-\infty)$ contributes an uncertainty of order $1/g$ to the experimental result, while an uncertainty in $D$ causes an effect at order $g$.  A short computation then shows that minimizing  the total uncertainty over both $g$ and $\Delta_{unc} P_D$ yields a non-zero value.   So, unless the coefficient of $D$ in the $O(g^2)$ term can be made arbitrarily small at {\it fixed} $g$, there would be some best resolution beyond which no interaction of our form would be able to measure  $[B]_{c=\tau_0}$.

There are two standard ways to deal with this issue.  In non-relativistic quantum mechanics, where there is well-defined notion of absolute time, it is common to consider time-dependent interactions as in \cite{vonN}.  In that case one can show that the $O(g^2)$ effects become small in the limit where the interaction happens very quickly, so that the time dependence is effectively given by some $\delta (t-t_0)$. Although our interactions are not explicitly dependent on the parameter time $\lambda$,  we will use a similar argument here, showing that the $O(g^2)$ terms become negligible in the semi-classical limit of large clock momentum $p_c$, which in particular has the consequence that the interactions last only a short parameter time $\lambda$.   In contrast, the other standard solution to this issue is to explicitly modify the interaction at $O(g^2)$ in order to cancel the dangerous term.  Such modifications are referred to as ``compensating devices'' in \cite{BR1,BR2,DeWitt}.  In effect, the analysis below shows that no compensating devices are required for our experiment at large $P_c$.

To begin, we use (\ref{conjexp}) to calculate the second order corrections to (\ref{DPD}):

\begin{eqnarray}
\label{DPD2}
\Delta P_D &+& g [B]_{c=\tau_0} + O(g^3)
=
(ig)^2 \int_{{|\lambda_1| \ge |\lambda_2| } \atop {{\rm sign} (\lambda_1) = {\rm sign}(\lambda_2)}} d\lambda_1 d\lambda_2 {\rm sign}\ [ (\lambda_1)] \ [ \Delta H_0(\lambda_2), [\Delta H_0(\lambda_1), P_D]]   \cr
&=&
i(ig)^2 \int_{{|\lambda_1| \ge |\lambda_2| } \atop {{\rm sign} (\lambda_1) = {\rm sign}(\lambda_2)}} d\lambda_1 d\lambda_2 [\rm{sign}(\lambda_1)] \ D_0(\lambda_2) [ (F_{\tau_0} B)_0(\lambda_2), (F_{\tau_0} B)_0(\lambda_1)] . \ \ \
\end{eqnarray}
In the final step we have used that, when evolved with the free Hamiltonian $H_0$, the operators $D$ and $F_{\tau_0}B$ commute at any pair of times.  The size of this integral is governed by properties of the commutator
\begin{eqnarray}
\label{lcom}
&[ (F_{\tau_0} B)_0(\lambda_2), (F_{\tau_0} B)_0(\lambda_1)] &= \cr & F_{\tau_0,0} (\lambda_1) F_{\tau_0,0} (\lambda_2) [B_0(\lambda_2), B_0(\lambda_1)]& +   [ F_{\tau_0,0} (\lambda_2), F_{\tau_0,0} (\lambda_1)]  B_0(\lambda_2) B_0(\lambda_1) .
\end{eqnarray}

We will consider each term in (\ref{lcom}) separately.  We begin with the first term.  Recall that $F_{\tau_0,0} (\lambda)$ vanishes as $\lambda \rightarrow \pm \infty$ and and integrates to $\openone$, so that it acts like a somewhat smoothed version of a Dirac delta function $\delta (\lambda - \lambda_0)$, where $\lambda_0$ is determined by the state to which the operator is applied; i.e., it acts something like $\frac{1 }{\sqrt{2\pi} \Delta \lambda}  e^{-(\lambda - \lambda_0)^2/2 \Delta \lambda^2}$ for some width $\Delta \lambda \sim \frac{ m_c \Delta c}{p_c}$, where $\sim$ denotes approximate equality in the semi-classical limit, $p_c$ is determined by the state, and $\Delta c$ is the width of the function $f'(c-\tau_0)$ used to define $F_{\tau_0}$.  Taking $p_c$ large,  $\Delta \lambda$ is small and it is useful to expand the commutator
$[B_0(\lambda_2), B_0(\lambda_1)]$ in powers of $(\lambda_1 - \lambda_2)$.  Noting that the commutator vanishes for $\lambda_1 = \lambda_2$, we may write:
\begin{equation}
[B_0(\lambda_2), B_0(\lambda_1)] = (\lambda_1 - \lambda_2) C(\lambda_1) + O([\lambda_1 - \lambda_2]^2),
\end{equation}
for some operator $C(\lambda_1)$.
As a result, in the limit of small $\Delta \lambda$, the first term in (\ref{lcom}) yields a term in (\ref{DPD2}) of order $\Delta \lambda \sim  \frac{m_c \Delta c}{p_c}$. which in particular vanishes in the limit of large $p_c$.

To deal with the second term in (\ref{lcom}), recall that $F_{\tau_0,0}(\lambda_i) =  \frac{\partial}{\partial \lambda_i} f(c(\lambda_i) - \tau_0)$, so that the commutator in the second term is
\begin{equation}
\label{Ftcom}
[F_{\tau_0,0} (\lambda_2), F_{\tau_0,0} (\lambda_1)] =
\frac{\partial}{\partial \lambda_2}  \frac{\partial}{\partial \lambda_1} [ f(c(\lambda_2) - \tau_0), f(c(\lambda_1) - \tau_0) ].
\end{equation}
Since we wish to consider the large $p_c$ limit where the clock behaves semi-classically, we
may approximate the commutator $[ f(c(\lambda_2) - \tau_0), f(c(\lambda_1) - \tau_0) ]$ as
\begin{eqnarray}
\label{fcapprox}
[ f(c(\lambda_2) - \tau_0), f(c(\lambda_1) - \tau_0) ] &\sim& f'(c(\lambda_2) - \tau_0)f'(c(\lambda_1) - \tau_0) [c(\lambda_2, c(\lambda_1)]  \cr &=& f'(c(\lambda_2) - \tau_0)f'(c(\lambda_1) - \tau_0) \frac{\lambda_2 - \lambda_1}{m_c},
\end{eqnarray}
where one may check that the corrections are higher order in $\frac{\hbar}{p_c \ \Delta c}$.

We now insert (\ref{fcapprox}) into (\ref{DPD2}) and integrate by parts to move both derivatives onto the $B_0(\lambda_i)$ factors.   In doing so,  boundary terms at $\lambda_{1,2} \rightarrow \pm \infty$ vanish since any derivative $ \frac{\partial^n}{\partial c^n} f(c-\tau_0)$ vanishes rapidly as $\lambda \rightarrow \pm \infty$.  Boundary terms at $\lambda_1 = \lambda_2$ vanish because (\ref{fcapprox}) contains an explicit factor of $\lambda_1 - \lambda_2$.     The result is a term of the form
\begin{eqnarray}
\label{ibp}
&i(ig)^2& \int_{{|\lambda_1| \ge |\lambda_2| } \atop {{\rm sign} (\lambda_1) = {\rm sign}(\lambda_2)}} d\lambda_1 d\lambda_2  \ [{\rm sign} (\lambda_1)] \ \cr
&\times&  f'(c(\lambda_2) - \tau_0)f'(c(\lambda_1) - \tau_0) \frac{\lambda_2 - \lambda_1}{m_c} \frac{\partial^2}{\partial \lambda_1 \partial \lambda_2} \Bigl( D_0(\lambda_2) B(\lambda_2) B(\lambda_1) \Bigr)  ,
\end{eqnarray}
together with a boundary term at $\lambda_2=0$.  Again taking the width of $f'$ to be $\Delta c$, in the  limit of large $p_c$ (with fixed behavior for the other degrees of freedom that define $D, B$) the contribution (\ref{ibp}) is of order $1/ p_c^3$.  The boundary term at $\lambda_2 =0$ is similar to (\ref{ibp}) without the integral over $\lambda_2$ and setting $\lambda_2=0$.  As a result, it is of order $1/p_c^2$.

We conclude that the second order contribution to (\ref{DPD}) can be made arbitrarily small compared to $g [B]_{c=\tau_0}$, even at fixed $g$, by taking $p_c$ large.   As a result, our measurement of $[B]_{c=\tau_0}$ can be made arbitrarily accurate in the limit of large $p_c$ and small $g$.

\section{Discussion}

\label{disc}

The goal of this work was to clarify the interpretation of certain evolving constants of motion $[B]_{c=\tau}$ by showing that a certain class of detectors, governed by appropriate interactions and prepared in appropriate initial states, can be said to measure such observables.  In particular, we showed that adding a suitable interaction $g \Delta H$ to a Hamiltonian constraint of the form (\ref{HC0}) results in perfect correlations between $[B]_{c = \tau}$ and the detector observable $P_{D,g}(+\infty) - P_{D,g}(-\infty)  = [P_D]_{c = +\infty} -  [P_D]_{c = - \infty} $  in the limit of small coupling constant $g$.  We have also shown that, for large clock momentum $P_c$, one may choose states sharply peaked on the spectrum of $P_{D,g}(-\infty)$ while still neglecting $O(g^2)$ corrections so that this effectively becomes a correlation between $[B]_{c = \tau}$
and $P_{D,g}(+\infty)  =  [P_D]_{c = + \infty} $.  This is the desired result.  In analogy with \cite{vonN}, it is natural to call such a situation a measurement of the observable  $[B]_{c = \tau}$.

Our discussion had much in common with classic treatments of measurement theory \cite{vonN,BR1,BR2,DeWitt}, and our study of evolving constants at large $P_c$ is closely related to that of \cite{BDT}. Indeed, the flavor of the results of \cite{BDT} is that at large clock momentum the operators $[B]_{c = \tau}$ behave much like the familiar Heisenberg picture operators of non-relativistic quantum mechanics with an external time.  (One may think of \cite{BDT} as re-coding classic results of \cite{BD1,TB,JBH,JJH,CK1,CK2} concerning the WKB approximation in the language of evolving constants of motion.) In this sense, it is no surprise that $[B]_{c=\tau}$ can be measured by an appropriate interaction.  The technical details added here concern the perturbation theory associated with our interaction $g \Delta H$ (as interactions between the clock $c$ and other degrees of freedom were not allowed in \cite{BDT}) and an explicit discussion of measurement issues.   Our approach also has much in common with \cite{CD}.

While our analysis made essential use of the large $P_c$ limit,
we find it plausible that high resolution measurements of an evolving constant $[B]_{c = \tau}$ can also be attained with $P_c$ fixed.  
  As noted in section \ref{meas}, a standard way to achieve high resolution measurements is to introduce additional explicit $O(g^2)$ compensation terms \cite{BR1,BR2,DeWitt}  in the interaction which cancel the $O(g^2)$ effect generated by our interaction $g \Delta H$.  The point is that since there are few constraints on the allowed $O(g^2)$ interactions in our minisuperspace context, one expects that a sufficiently complicated term would indeed cancel the $O(g^2)$ effect generated by $g \Delta H$.  

It would very interesting to explore such compensation terms in detail.  After all, in a real gravitational system, there would be some constraint on the total energy that our clock could have (say, given its size) without collapsing into a black hole.  Thus, if one {\it cannot} obtain arbitrarily high resolution at fixed clock momentum $P_c$, one might expect gravity to impose some limitation on the resolution of measurements of observables\footnote{Note that here we discuss the resolution to which we can measure the value of given observable, which may or may not be localized in space.  This is in contrast with the discussion of e.g. \cite{GMH} or \cite{GPP1,GPP2} concerning the degree to which a given observable can be localized in position space.}.     Such limitations might then have interesting implications for the structure of observables in a full quantum theory of gravity.

We should also comment on the particular form (\ref{Bct}) of the operators $[B]_{c =\tau}$ measured by our setup.  Such operators were described by choosing a particular operator $c$ to act as a clock, taking $c,P_c$ operator to appear in $[B]_{c =\tau}$ in a certain combination, and by assuming certain properties for this operator.   However, the result (\ref{DPD}) describing first order correlations between 
$[B]_{c =\tau}$ and $\Delta P_D$ is extremely robust, and is independent of most of these considerations.  Thus, as the reader may check, this particular result generalizes immediately to measurements of any operator of the form $\int d \lambda e^{iH\lambda} Ae ^{-iH\lambda}$ so long as $A$ commutes with the detector momentum $P_D$.  In particular, this result would apply to the case where $A$ is simply a projector onto some region of the configuration space as advocated in \cite{partial,CRbook,Dolby}. It would be interesting to extend the other results above to this context, and to see if this can be done in a way that does not require treating the clock variable differently from the other degrees of freedom\footnote{\label{f2}See \cite{CD} for a study of such measurements that achieves this symmetry, but which derives a somewhat different set of results.  In particular, \cite{CD} uses a different notion of time evolution and does not analyze the 2nd order terms.} .  

Finally, we note that additional issues will arise at the level of quantum field theory.  Due to the presence of arbitrarily high frequency modes, making the duration of the experiment short does not necessarily lead to small $O(g^2)$ effects.  Instead, it can merely transfer them to these high-frequency modes.   One therefore expects that explicit compensation terms or the equivalent must be introduced to produce high resolution experiments, and one must ask whether this can be done in a manner consistent with locality.  Some steps toward answering these questions will be explored in \cite{GMtoappear}.

\appendix

\section{Clocks are free for large $P_c$}

\label{freeclock}

This appendix gives a short proof of the following useful technical result: in the limit where the auxiliary state $\aux \psi \rangle$ is supported at large clock momentum $P_c$ (with the state of the other degrees of freedom held fixed), we may write

\begin{equation}
\label{freef}
e^{iH_g \lambda} F_\tau e^{-iH_g \lambda} \aux \psi \rangle
= e^{iH_0 \lambda} F_\tau e^{-iH_0 \lambda} \aux \psi \rangle  + O(1/p_c),
\end{equation}
so that the interaction does not affect the clock degree of freedom at large $p_c$.
Here $p_c$ is a typical eigenvalue of $P_c$ associated with the state $\aux \psi \rangle$.

This result may be motivated by noting that the clock should behave semi-classically at large $p_c$, and that the interaction effectively describes a potential for the clock which depends on the other degrees of freedom.  Taking $p_c$ large with the other degrees of freedom held fixed, we expect the behavior to be that of a classical particle with large kinetic energy moving over a small potential $V(c)$ of fixed extent.  If $V(c)$ vanishes at infinity, then, in perturbation theory, by the time it reaches $c$ the the particle receives an impulse $\int V'(c(t)) dt \sim \frac{m}{p_c} \int^c V'(\tilde \tau) d \tilde \tau = \frac{mV(c)}{p_c}$ which vanishes in the limit of large $p_c$.  The effect of the potential on $c(t)$ is proportional to the time integral of this impulse.  As a result, it also vanishes in the limit of large $p_c$. Thus, in the quantum context one also expects the effect of the potential to vanish in the limit of large $p_c$.

The above argument is heuristic.  One way to make it rigorous is to work through the details of the multi-variable semi-classical approximation.  But one may also use the perturbation expansion (\ref{Uexp}) to reduce the calculation to a one-dimensional problem (which is then easily evaluated via familar WKB techniques).  To do so, write
$\Delta H = F_{\tau_0} \widetilde{\Delta H}$, $H_0 = \frac{1}{2m_c} P_c^2 + \tilde H_0$, $\lambda_2 := \lambda - \lambda_1$ and thus

\begin{eqnarray}
&e^{iH_g \lambda}& F_\tau e^{iH_g \lambda} \aux \psi \rangle
-  e^{iH_0 \lambda} F_\tau e^{iH_0 \lambda} \aux \psi \rangle \cr
&=&ig \int_0^\lambda d \lambda_1 \left( e^{iH_0 \lambda_1} F_{\tau_0} \widetilde{\Delta H} e^{iH_0 \lambda_2}  F_\tau e^{-iH_0 \lambda} -  e^{iH_0 \lambda} F_\tau e^{-iH_0 \lambda_2}  F_{\tau_0} \widetilde{\Delta H}  e^{-iH_0 \lambda_1}\right)
\aux \psi \rangle  + O(g^2)  \ \ \ \ \ \ \ \cr
 \cr
&=& ig\int_0^\lambda d \lambda_1 \left( e^{i\frac{P_c^2}{2m_c} \lambda_1} F_{\tau_0}  e^{i\frac{P_c^2}{2m_c} \lambda_2}  F_\tau e^{-i\frac{P_c^2}{2m_c} \lambda} -  e^{i\frac{P_c^2}{2m_c} \lambda} F_\tau e^{-i \frac{P_c^2}{2m_c} \lambda_2}  F_{\tau_0} e^{-i \frac{P_c^2}{2m_c} \lambda_1}\right)\cr &\times& e^{i \tilde H_0 \lambda_1} \widetilde{\Delta H} e^{-i \tilde H_0 \lambda_1}
\aux \psi \rangle  + O(g^2) . \ \ \ \ \ \ \ 
\end{eqnarray}

Now, much in section \ref{pert}, at large clock momentum $p_c$ the operators $F_{\tau_0}, F_{\tau}$ become sharply peaked functions of $\lambda$ when acting on states that are sharply peaked in $c$.  Consider such a state $\aux \psi \rangle$. Then, in this limit, the integrand is very small unless $\lambda_2$ takes some particular value $\bar \lambda_2$ required for the peak of the wavefunction to advance from $\tau$ to $\tau_0$.  I.e., the integrand has support only for $\lambda_1 \approx \lambda - \bar \lambda_2$.  We may therefore write

\begin{eqnarray}
\label{long}
&e^{iH_g \lambda}& F_\tau e^{iH_g \lambda} \aux \psi \rangle
-  e^{iH_0 \lambda} F_\tau e^{iH_0 \lambda} \aux \psi \rangle \cr
&=& ig\int_0^\lambda d \lambda_1 \left( e^{i\frac{P_c^2}{2m_c} \lambda_1} F_{\tau_0}  e^{i\frac{P_c^2}{2m_c} \lambda_2}  F_\tau e^{-i\frac{P_c^2}{2m_c} \lambda} -  e^{i\frac{P_c^2}{2m_c} \lambda} F_\tau e^{-i \frac{P_c^2}{2m_c} \lambda_2}  F_{\tau_0} e^{-i \frac{P_c^2}{2m_c} \lambda_1}\right)\cr &\times& e^{i \tilde H_0 (\lambda - \bar \lambda_2)} \widetilde{\Delta H} e^{-i \tilde H_0 (\lambda - \bar \lambda_2)}
\aux \psi \rangle  + O(g^2)  \ \ \ \ \ \ \ \cr
&\approx& \left( e^{i\left(\frac{P_c^2}{2m_c} +g F_{\tau_0} \right) \lambda}   F_\tau e^{i\left(\frac{P_c^2}{2m_c} +g F_{\tau_0}\right) \lambda} - e^{i\frac{P_c^2}{2m_c} \lambda} F_\tau e^{-i\frac{P_c^2}{2m_c} \lambda} \right) \cr &\times& e^{i \tilde H_0 (\lambda - \bar \lambda_2)} \widetilde{\Delta H} e^{-i \tilde H_0 (\lambda - \bar \lambda_2)}
\aux \psi \rangle  + O(g^2).  \ \ \ \ \ \ \
\end{eqnarray}
In the final step we have used the fact that the $\widetilde {\Delta H}$ factor does not participate in the integral in the 2nd line to use (\ref{Uexp}) in reverse, thinking of $gF_{\tau_0}$ as a perturbation on the free one-dimensional Hamiltonian $\frac{1}{2m_c}P_c^2$.  The point is that the term in parentheses on the final line depends only on $c, P_c$.  Thus, one may calculate its matrix elements using the one-dimensional WKB approximation.  The WKB approximation for the perturbed term involves expressions of the form
\begin{equation}
\int dc \sqrt{p_c^2 - 2g m_c F_{\tau_0}} = \int dc \ p_c (1 - \frac{g m_c F_{\tau_0}} {p_c^2} + \dots).
\end{equation}
Since $F_{\tau_0}$ is of order $p_c$, we see that the second term is smaller than the first by a factor of order $1/p_c$.  Thus, the entire right-hand-side of (\ref{long}) is of order $1/p_c$. The terms at higher order in $g$ can be dealt with in a similar way.  Each term factorizes into a clock piece involving only $c,P_c$, and a piece involving the other degrees of freedom.  Each clock piece can then be identified with a term in the expansion of $\left( e^{i\left(\frac{P_c^2}{2m_c} +g F_{\tau_0}\right) \lambda}   F_\tau e^{i\left(\frac{P_c^2}{2m_c} +g F_{\tau_0} \right) \lambda} - e^{i\frac{P_c^2}{2m_c} \lambda} F_\tau e^{-i\frac{P_c^2}{2m_c} \lambda} \right)$, and so is again of order $1/p_c$ as above.  Since the states used above (localized packets in $c$) span the full Hilbert space we conclude that (\ref{freef}) holds, at least when treated perturbatively in $g$.

\begin{acknowledgments}  The author thanks Bianca Dittrich, Rodolfo Gambini, Rafael Porto, Jorge Pullin for interesting discussions on measurement theory and evolving constants of motion.  He especially thanks Karel Kucha\v{r} for discussions of such observables over many years, and Steve Giddings for more recent discussions emphasizing the importance of relating observables  to measurements.   The author is also indebted to Bryce DeWitt for many discussions of such issues during the author's time as a graduate student. This work was supported in part by the US National Science Foundation under Grant No.~PHY05-55669, and by funds from the
University of California.
\end{acknowledgments}

\end{document}